# The Orbit of Planet Nine Derived from Engineering Physics

R. Finch[1], M. A. Galiazzo[2]

## Abstract
Several papers have recently suggested the possible presence of a ninth planet (Planet X) that might explain the gravitational perturbations and orbital arrangement of a number of detached Trans-Neptunian objects. To analyze the possibility further, we have applied celestial mechanics, engineering physics and statistical analysis to develop improved estimates of the planet's primary orbital elements and mass from first engineering principles, using the orbital characteristics of both the original group of 6 objects analyzed and also a second group comprising the original 6 together with 6 additional long-period asteroids selected by the authors. We show that the driving force behind the observed clustering is gravitational torque that arranges the orbits of asteroids in a systematic, orderly manner, and we develop the associated equations of motion. As evidence we show that the expected effects are fully apparent in the orbital characteristics of the correlated bodies involved, including most strikingly regarding their orbital planes, azimuth orientations and specific relative angular momenta, which we show generates a highly unexpected form of resonance in their relative angular momenta. We further show that the coordinates of Planet Nine's orbit are close to the original values proposed recently by other authors, although we prove that its period has to be dramatically smaller than that proposed in recent literature (Batygin & Brown 2019) at about 3500 yrs, the eccentricity is near 0.65, and its mass approximately 8.4 times the Earth's mass. Given the predicted orbit we show that the planet has apparently created numerous mean motion resonances, of which seven are noted specifically. The overall analytical predictions are shown to be supported by the statistical characteristics of the asteroid data across both data samples and by a proof-of-concept orbital simulation of Planet X's gravitational effect on the motion of two dwarf planets integrated over a 40 million year time frame.

## 1. Introduction

"Planet Nine" is the recently coined term for the previously designated "Planet X", a hypothetical planet believed for over two centuries to exist somewhere in the far reaches of the solar system that might explain orbital irregularities of Neptune and Uranus. In the 1980s extensive effort to research and locate the proposed planet was undertaken by the Naval Observatory under the direction of Robert Harrington (Harrington 1988), who even used the Pioneer probes 10 and 11 to help with the search. Interest in the subject stalled following his death in 1993, but returned the next decade with a number of related studies (Collander-Brown et al. 2000; Brunini & Melita 2002; Melita et al. 2003, 2004; Matese et al. 2005; Gomes et al. 2006; Lykawka & Mukai 2008). Of particular significance was a 2014 article in *Nature* by Gemini Observatory astronomers Chad Trujillo and Scott Sheppard announcing their discovery of the long-period dwarf planet 2012 VP113 (Trujillo & Sheppard 2014). That body provided evidence that the discoveries of two large–mass, long–period bodies, the first being the dwarf planet Sedna discovered in 2003, might be indications of another much larger planetary body orbiting in the outermost solar system. In February 2016 two scientists at the California Institute of Technology proposed that such a planet might explain the unusual orbital configuration of a particular group of 6 long-period, Trans-Neptunian objects, which included the above two dwarf planets. Astronomers Michael Brown, one of Sedna's discoverers, and Konstantin Batygin, predicted that a massive outer planet should be the most likely explanation for the unusual arrangement of those orbits and that the planet was apparently on or near a long elliptical orbit that they estimated using an elaborate simulation (Batygin & Brown 2016). They further proposed that the unseen planet has an approximate mass of about 10 times that of Earth and a period in the range of 15,000 years, although the researchers have subsequently amended these numbers in a later publication (2019). Since then a number of other research efforts have extended the Caltech research with their own simulations using additional reference objects and related extensions of their original model (de la Fuente Marcos 2014; Malhotra et al 2016; Gomes et al 2016). All have come to similar conclusions that at least one extremely large object exists somewhere in the outer solar system. The results herein provide additional support for the existence of Planet X, and significantly improves the fidelity of recently published orbital parameters.

| | | |
|---|---|---|
| Mattia A. Galiazzo | mattia.galiazzo @univie.ac.at | 436-812055-0608 |
| Robert Finch | rfinch15@stny.rr.com | 607-754-2630 |

[1] Senior Engineer ( Retired ), Lockheed Martin Corporation, Mission Systems and Sensors,1801 NY-17C, Owego, New York
[2] Department of Astrophysics, University of Vienna, Turkenschantzstrasse 17, 1180, Vienna, Austria

The basic premise of our approach is that the orbits of selected trans-Neptunian bodies, hereafter simply referred to as "asteroids", are being systematically arranged by the undetected Planet X (PX). The asteroids' orbits may appear to be clustered in a somewhat arbitrary manner, but in fact they actually follow a regular pattern in relation to PX's orbit, which becomes ever more obvious as the number of affected asteroids increases. The basic issues for analysis are whether the pattern can be deciphered analytically using the orbits of the asteroids making up the pattern, and if so how the proposed model and associated description can be used in estimating the orbit of Planet X. We show in the following discussion that both issues can be explained with standard dynamics to estimate PX's orbit using Batygin's and Brown's (2016) original 6 objects and a second estimate that uses the original 6 plus 6 additional asteroids selected by the authors.

The goal of the analysis focuses on developing a set of estimates of selected parameters of Planet X's orbit using the orbital characteristics and interrelationships among the two sets of reference asteroids indicated, engineering physics and standard statistics as described in standard texts (Moulton 1984; Hildebrand 1962). The estimates comprise a set of mathematical models combined with the statistical characteristics of the reference asteroids' orbits and mass being used. The analysis follows directly from two fundamental facts of orbital mechanics: (1) the Sun together with everything orbiting it is a closed system, which implies that the angular momentum of the system is conserved; (2) there are effectively no torques acting on the system except those arising from within the system itself, which implies the net torque on the system must sum to zero. Using these two laws and the orbital characteristics of the reference asteroids, the following analysis develops closed-form, analytical estimates for the inclination, longitude of the ascending node and argument of perihelion of the major axis of Planet X, its specific relative angular momentum, semi major axis length, eccentricity, period, perihelion and mass. The angular coordinates and mass were found to be in rough agreement with values reported in recent literature, but the semi major axis length and orbital period are both shown to be dramatically smaller.

## 2. Methods

The central claim of the analysis is that PX induces gravitational torque on asteroids orbiting the Sun in a privileged region of influence, which gradually arranges their orbits over time in a systematic, regular manner. The dominant result of this effect is that the torque induced by PX drives the orbital plane of each affected asteroid toward aligning with its own, which in turn causes the orientation of the angular momentum vector of the asteroid to precess towards that of PX as well. A secondary result of the torque is that the major axis of each affected asteroid shifts its spatial orientation in azimuth towards aligning with either PX's approach or retreat path, whichever is closer in angular separation. Over time these two effects create a distinct pattern in the orbits of affected asteroids, and this pattern can be used to discover a number of orbital characteristics about the unseen object responsible.

To understand how all this is happening, when we consider just one of the 6 asteroids, two primary forces pull the object in different directions: the gravitational force $\boldsymbol{F_i}$ pulls it towards the Sun and $\boldsymbol{F_{pi}}$ towards PX. In general these two forces affect the asteroids along different lines of action, which cause a torque to develop, the dominant effect in this case being the torque induced by PX on each of the reference asteroids outside its orbital plane (see Figure 1 below). To understand how this works in our solar system, we formally compute the torque $\boldsymbol{\tau_i}$ on an object due to the influence of PX. If $\boldsymbol{F_i}$ and $\boldsymbol{F_{pi}}$ represent the forces of gravity between asteroid i and the Sun and PX, respectively, and $\boldsymbol{r_{is}}$ the vector distance from the Sun to asteroid i, then the induced torque $\boldsymbol{\tau_i}$ is computed as

$$\boldsymbol{\tau_i} = \boldsymbol{r_{is}} \times (\boldsymbol{F_{pi}} + \boldsymbol{F_i}) \quad (1)$$

$$= \boldsymbol{r_{is}} \times \boldsymbol{F_{pi}}, \quad (2)$$

where the simplification occurs because the cross product of parallel vectors is always zero. The angle between Planet X's gravitational force acting on the reference object and its distance to PX correlates with the fraction of the force causing the torque. Using the following diagram we can show how this process occurs dynamically.

Figure 1 below depicts the orbits of PX and a reference asteroid under the influence of PX's attraction. The bold line represents an edge-on view of PX's orbital plane with the orbit's major axis in the plane of the illustration. The ellipse shows an oblique view of the orbit of an in-coming asteroid with its out-going path tilted downward from the reader's perspective looking downward on the asteroid's orbital plane. The angle between the two planes is equal to the angle between the two angular momentum vectors $\boldsymbol{L_p}$ and $\boldsymbol{L_i}$, which are each normal to their respective orbital plane. However much these values change over time, the total angular momentum of the system $\boldsymbol{L_T}$ remains constant due to conservation of angular momentum. A theorem related to this law states that the net torque induced on the system is equal to the time rate of change of the system's total angular momentum. In particular, when there are no external forces acting on the system, the net torque on the system is zero, which implies that the net angular momentum is conserved and therefore constant and the system as a whole is not rotating. For the basic model we have presumed that the mass of the inner solar system is concentrated at the Sun, and that there are negligible interactions among the reference asteroids themselves or with other bodies in the solar system except of course for PX and the Sun, discussed in Section 2 below. These ideas are expressed in the following two equations

$$\boldsymbol{L}_T = \boldsymbol{L}_p + \sum \boldsymbol{L}_i \equiv \text{constant} . \qquad (3)$$

$$\boldsymbol{\tau}_p + \sum \boldsymbol{\tau}_i = \mathrm{d}\,\boldsymbol{L}_T / \mathrm{dt} = 0 , \qquad (4)$$

The primary, long-term effect of the torque is to drive the spatial orientation of each reference asteroid's angular momentum vector and orbital plane toward aligning with the planet's own. Using the following diagram we can see how this occurs conceptually. In the stable orbit of a single body, PX for example, moving under the influence of a second much larger body, namely the Sun, the planet's angular momentum vector is normal to its orbital plane and constant in inertial space over time. If a third body, say asteroid i, with a much smaller mass than either of the other two bodies is sufficiently close to PX, then per equation (2) a torque arises normal to the plane passing through the three bodies shown, whose action can be interpreted using the well-known 'right-hand rule'. To apply this rule to the example below, the reader should place the fingers of the right hand to point along the range vector from the Sun to asteroid i, and bend the fingers to point to Planet X, the source of the torque. The thumb then points upwards in the direction of the torque vector, which is also the direction of the asteroid's orbital precession at each point in its orbit. Using this procedure the reader can easily infer that this upward force is present for most of the asteroid's orbit shown in the figure. The torque does go to zero, however, when the three objects line up and then reverses direction to pointing downwards when the asteroid is above PX's orbital plane near perihelion on the asteroid's semi major axis. But by rotating the higher side downward in this region, the oppositely directed force on the asteroid is also helping to drive the object's orbital plane toward PX's. The net effect of this process is that the entire orbit precesses in a way that causes the object's angular momentum vector and orbital plane to rotate slowly towards those of PX. All this said, however, there is a complication involved.

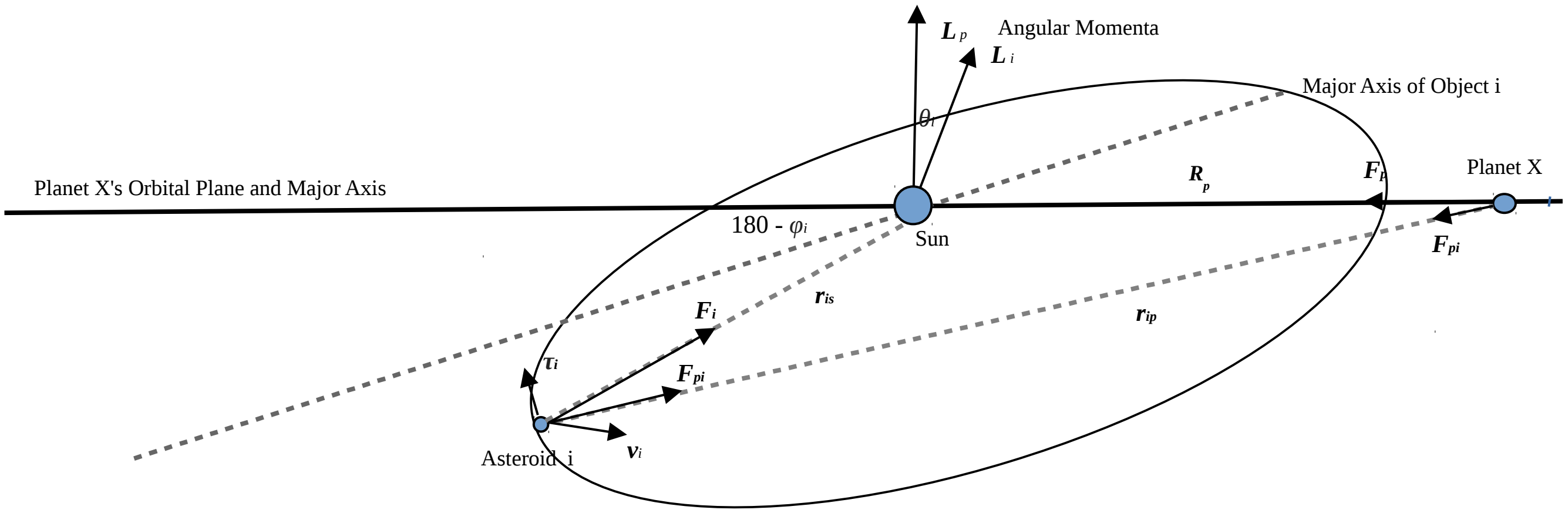

Fig 1 Planet X and an asteroid orbiting under its gravitational influence with the applicable forces, angular momenta and torque vectors indicated

The gravitational torque precession works as described above but varies in direction over the orbit. As a result there are regions in every orbit in which the torque is being applied in the "wrong" direction to shift the asteroid's orbit toward the planet's orbital plane. In Figure 1 if the asteroid's motion is backed up by about 20°, according to the right hand rule the torque is still being directed upward when downward is the "needed" direction at that point. This generally causes considerable oscillation near the region of convergence, and for low mass asteroids can even cause the inclination to diverge considerably, an effect possibly seen with a number of small high-inclination, low mass asteroids that have apparently fallen victim to this effect.

This intuitive argument can be described rigorously by leveraging a standard theorem from rotational dynamics, which states that for any vector function $\boldsymbol{f}$ rotating in a vector field at a time-varying rate of change $\boldsymbol{\xi}$, the vector's total rate of change is described as

$$d\,\boldsymbol{f} / dt = \boldsymbol{\xi} \times \boldsymbol{f} .$$

This relation holds in particular for the specific relative angular momentum $\boldsymbol{h}$ of a body with constant mass moving in a closed system under a central gravitational force:

$$d\,\boldsymbol{h} / dt = \boldsymbol{\xi} \times \boldsymbol{h} . \qquad (5)$$

When we consider the relationship of the specific relative angular momentum of PX and an arbitrary asteroid i, $\boldsymbol{h}_p$ and $\boldsymbol{h}_i$ , respectively, the difference of these vectors rotates with angular velocity $\boldsymbol{\xi}_i$ about the *z* axis of the angular momentum $\boldsymbol{h}_T$, and from the above discussion we have

$$d\,( \boldsymbol{h}_i - \boldsymbol{h}_p ) / dt = \boldsymbol{\xi}_i \times ( \boldsymbol{h}_i - \boldsymbol{h}_p ).$$

Because the mass of PX is estimated to be several orders of magnitude larger than that of any one or all of the asteroids combined (see below), its angular momentum is very nearly constant and we can take its derivative to be negligible in relation to that of an asteroid, which then yields the asteroids' approximate equations of motion (discussed further in Section 2.2):

$$d\,\boldsymbol{h}_i / dt = \boldsymbol{\xi}_i \times ( \boldsymbol{h}_i - \boldsymbol{h}_p ). \qquad (6)$$

Two assumptions given at the beginning of this section regarding the system's mass distribution and gravitational independence of the system's various bodies need some further discussion. Although Planet X and the reference asteroids will be shown to spend their orbital journeys predominantly far apart on opposite sides of the Sun, on occasion when an asteroid and the planet are both near perihelion in the same time frame, the two bodies can come within 20 AU or so of each other, and when that happens a substantial perturbation of the asteroid's orbit will almost certainly occur. There is a mitigating factor, however. Whenever an asteroid is orbiting inside the orbit of a much larger body having a near circular orbit, an effect called the Kozai mechanism occurs when the two bodies come close to one another that preserves the asteroid's *z* component of angular momentum normal to the perturbing body's orbital plane. Even though the encounter can radically affect the *x* and *y* components as the asteroid's inclination and eccentricity are shifted, the *z* component remains nearly constant. This configuration occurs approximately whenever an asteroid and PX are both near the Sun at about the same time frame, since the asteroid does stay inside PX's orbit near the Sun (shown below), and its distance from the Sun varies only moderately in this portion of its orbit, therefore approximating a circular orbit near the Sun. The Kozai conditions are therefore approximately satisfied near perihelion, the asteroid's *z* component of angular momentum remains nearly constant and the asteroid's orbital planar orientation remains largely the same following the close encounter.

In another scenario of concern when an asteroid nearing its perihelion comes too close to either of the large outer planets Uranus and Neptune, the passage is most likely to be on the outer side of the planets' orbit and the Kozai mechanism is not applicative. What specifically can happen in this situation in general is unpredictable a priori: either the asteroid remains somewhere within the planet's region of influence and continues as one of the asteroids being rearranged or else the body is ejected from the region and becomes affected by the system's broader dynamics in unpredictable ways. Whatever such ejected bodies do is beyond the scope of this study, which of course is specifically to use the orbits of a collection of asteroids whose orbits have been recently observed to follow a certain broad pattern of behavior that can be analyzed to infer the presence and orbital characteristics of the hypothetical Planet X.

### 2.1 Orbital Plane and Azimuth Orientation

The foundation of the model rests on the estimate of the spatial coordinates of Planet X's major axis. If the orbits of the reference asteroids are in fact being systematically arranged by PX as proposed, then the orientation of their angular momentum vectors orbital planes in space, should be at least moderately close to that of PX. We can "test" this theory by using the planes of the asteroids' orbits to estimate PX's orbital plane statistically and then using this plane to assess how well its orientation compares statistically with those of the reference asteroids. To do this we first calculate the unit vectors normal to each of the reference asteroid's orbital plane by using each body's standard orbital coordinates: inclination ( $\iota$ ), longitude of the ascending node ($\Omega$ ) and angle of perihelion ($\omega$ ). The desired unit vector is computed using the following coordinate transformation matrix $\boldsymbol{U}$ that transforms a Cartesion 3x1 vector in the heliocentric equatorial coordinate system into the body's local orbital coordinate system. The inverse of $\boldsymbol{U}$, also its transpose $\boldsymbol{U}^T$, performs the inverse transformation from the body's local coordinate system to the heliocentric equatorial system.

$$\boldsymbol{U} = \begin{matrix} \cos\Omega\,\cos\omega - \sin\Omega\,\cos\iota\,\sin\omega & \sin\Omega\,\cos\omega + \cos\Omega\,\cos\iota\,\sin\omega & \sin\iota\,\sin\omega \\ -\cos\Omega\,\sin\omega - \sin\Omega\,\cos\iota\,\cos\omega & -\sin\Omega\,\sin\omega + \cos\Omega\,\cos\iota\,\cos\omega & \sin\iota\,\cos\omega \\ \sin\iota\,\sin\Omega & -\sin\iota\,\cos\Omega & \cos\iota \end{matrix}$$

In the coordinate system of the orbital plane, the unit vector normal to that plane that points along the orbit's major axis toward perihelion is simply $(0, 0, 1)^T$. When this vector is transformed to the heliocentric equatorial system using $\boldsymbol{U}^T$ the result is simply the third row of the $\boldsymbol{U}$ matrix, a key vector used throughout the analysis that we refer to as the "normal unit vector", designated as $\boldsymbol{u}$. If the asteroids' orbits are being systematically driven to align with PX's orbital plane as we propose, then $\boldsymbol{u_p}$ should be strongly correlated with the normal unit vectors of all the clustered asteroids, and we can therefore use the asteroids' corresponding known unit vectors to estimate the value of $\boldsymbol{u}_p$ from this correlation. To do this we use the standard optimization formalism to find the value of $\boldsymbol{u_p}$ that minimizes the following objective function, where each term in the sum is the square (vector dot product) of the difference between the normal unit vectors of each asteroid i and the unknown normal unit vector of PX:

$$f(\boldsymbol{u}_p) = \Sigma_i\,(\boldsymbol{u_p} - \boldsymbol{u_i}) \cdot (\boldsymbol{u}_p - \boldsymbol{u_i})\,. \qquad (7)$$

To find the value of $\boldsymbol{u_p}$ that minimizes $f$, we take partial derivatives of $f$ with respect to each of the three components of $\mathbf{u_p}$, set each resulting derivative to zero, and solve for $\mathbf{u_p}$. The solution turns out to be the simple average, with $n$ being the number of asteroids used in the calculation:

$$\hat{\boldsymbol{u}}_p = \Sigma\,(1/n)\;\boldsymbol{u}_i\,. \qquad (8)$$

The resulting vector $\hat{\boldsymbol{u}}_p$ is typically not a unit vector, so the following normalization process is required:

$$\boldsymbol{u}_p = \hat{\boldsymbol{u}}_p \,/\, (\hat{\boldsymbol{u}}_p \cdot \hat{\boldsymbol{u}}_p)^{0.5}\,. \qquad (9)$$

From $\boldsymbol{u}_p$ the estimate of PX's inclination and longitude of the ascending node can be calculated directly using the normal unit vector's component elements, where the arc tangent calculation accommodates 360 degrees:

$$\Omega_p = \tan^{-1}\,[-\boldsymbol{u}_p(x) \,/\, \boldsymbol{u}_p(y)]\,. \qquad (10)$$

$$\iota_p = \tan^{-1}\;\boldsymbol{u}_p(z) \,/\, \sqrt{\boldsymbol{u}_p(x)^2 + \boldsymbol{u}_p(y)^2}$$

The angle between the planes of Planet X and a given reference asteroid i, which is equal to the angle between the two body's normal unit vectors, is then given by

$$\sin\,\theta_i = \|\,\boldsymbol{u}_p \times \boldsymbol{u}_i\,\|\,. \qquad (11)$$

Estimating the argument of perihelion is a little more involved. The dominant effect of PX's induced torque is to align the asteroids' orbital planes with that of Planet X as described above, but a smaller portion of that torque causes a shift in each asteroid's azimuth orientation as well. This torque can be inferred from Figure 1 in which $\boldsymbol{L}_p$ and $\boldsymbol{\tau}_i$ are slightly misaligned in such a way that $\boldsymbol{\tau}_i$ points primarily along the direction of PX's angular momentum $\boldsymbol{L}_p$ while leaving a small fraction of the torque vector directed along azimuth. Small though that effect may be, examination of the data clearly indicates that the effect has caused a double "attractor" direction to develop in azimuth. This dual effect is clearly evident when we examine the longitude of perihelion for each of the 12 asteroids given in column 12 of the table below: the lower cluster falls in the interval of about 355° to 400° and the upper of about 430° to 480°, which are expressed in angles from 360° to 540° throughout to make the clustering more numerically apparent. So, what does this double clustering mean?

The azimuth orientation of an asteroid's orbit is due to both the argument of perihelion and longitude of the ascending node, the sum being the longitude of perihelion $\varpi_p$, which is an approximate measure of the azimuth orientation of the major axis in inertial space. This is readily seen by examining the first row of the $\boldsymbol{U}$ matrix in which the first two components are approximately the cosine and sine of that longitude, both perturbed slightly by the cosine of inclination, which is not far from unity for most of the asteroids in our case. The longitude of the ascending node is primarily driven to align the asteroid's plane with that of Planet X as shown above, which leaves the argument of perihelion to respond to the smaller torque acting in the azimuth direction. The two attractors seen in the data turn out to be caused by the Planet X's approach and departure paths oriented 180° from the two attractor directions, which have their most apparent effect near perihelion where the asteroids come closest to the planet. This effect is discussed further in Section 2.3 below. The result is that the gravitational torque is forcing each asteroid's azimuth orientation to shift toward aligning with one or the other of PX's dominant orbital directions, whichever has the smallest angular separation from the asteroid's major axis. This effect is the key to estimating the argument of perihelion.

Given these dynamics, we would expect the major axis of Planet X to be oriented approximately at the mid point between the asteroids' two average azimuth distributions due to the symmetric nature of PX's orbital dynamics and geometry. The azimuth orientation of PX's orbit then should be that value of the longitude of perihelion that best reflects this azimuth symmetry, which is the value equidistant between the two average values of longitude of perihelion of the asteroids associated with each of the two paths, computed as

$$\varpi_{\text{p}} = 0.5 \left[ \Sigma_{\text{low}} (1/n)(\omega_i + \Omega_i) + \Sigma_{\text{high}} (1/m)(\omega_i + \Omega_i) \right] - 180°, \qquad (12)$$

where the two summations represent the averages of the lower and higher valued longitudes of perihelion, the number of asteroids included in each being n and m, respectively. The subtraction of 180° is due to the fact that Planet X and the reference asteroids have their perihelia on opposite sides of the Sun. Given that the planet's value of $\Omega_p$ has already been determined in (10) above, its argument of perihelion is therefore given by

$$\omega_p = \varpi_p - \Omega_p .$$

The angular deviation in azimuth of a given asteroid i from the average longitude of perihelion in the high or low cluster is therefore given by the following, where *k* is the total number of bodies in the particular cluster the asteroid is in:

$$\Delta \varpi_i = \left[ \Sigma_{\text{high/low}} (1/k)(\omega_i + \Omega_i) \right] - (\omega_i + \Omega_i)$$

Letting $\varphi_i$ be the angular separation between the major axes of Planet X and reference asteroid i, this angle is computed using the unit vector $\boldsymbol{w}$ in the orbital coordinate system that points in the direction of perihelion, which is defined by the first row of the $\boldsymbol{U}$ matrix. The separation angle $\varphi_i$ is then computed as

$$\sin \varphi_i = \| \boldsymbol{w}_p \times \boldsymbol{w}_i \| . \qquad (13)$$

Of the original 6 asteroids only two are in the high cluster, both being well on the high side within that cluster, when compared to the high cluster distribution with all 12 asteroids considered. This data bias of the smaller case produces an estimate of $\omega_p$ of about 153°, close to the value the Caltech astronomers proposed in their original paper based on the same 6 bodies, but this value is an artifact of the limited data sample as the analysis of all 12 asteroids clearly shows, which produces a much lower value of about 127°. The numerical results of these calculations are discussed in further detail in Section 3.0 and all results are summarized in the table below.

## 2.2 Specific Relative Angular Momentum

Estimating the remaining orbital elements of Planet X depends directly or indirectly on knowing its specific relative angular momentum $\boldsymbol{h}_p$, the angular momentum per unit mass, which is derived below using the system's total distribution of angular momentum. The estimation strategy uses a perturbation analysis that leverages the known angular momenta and spatial orientations of the reference asteroids and their relation to the angular momentum of the system as a whole. The following expressions state the angular momenta relationships for the system, the values of "*m*" being the mass associated with the individual body's angular momentum in the sum:

$$\boldsymbol{L}_T = \boldsymbol{L}_p + \Sigma \boldsymbol{L}_i \qquad (14)$$

$$= m_p \boldsymbol{h}_p + \Sigma m_i \boldsymbol{h}_i . \qquad (15)$$

When the first term on the right hand side of (15) is transferred to the left side, the asteroids' collective angular momentum can clearly be identified as a small perturbation $\delta \boldsymbol{L}_p$ from the angular momentum of Planet X without the asteroids' presence. The set of asteroids orbiting predominantly on the opposite side of the Sun from the planet changes the planet's orbit ever so slightly by minutely increasing the attractive force on the planet in the various directions of the asteroids' locations, which changes the angular momentum of the planet to keep the system's total constant over time. This on-going balancing process is the key to estimating the planet's specific relative angular momentum (SRAM).

When a planet orbiting the Sun has a collection of asteroids with only negligible mass orbiting it, the total angular momentum of the system $\boldsymbol{L}_T$ is essentially equal to that of the planet $\boldsymbol{L}_p$ . If the asteroids' mass are not negligible they add angular momentum to the system that the planet has to "balance" by changing its own to keep the total angular momentum of the system constant. The total angular momentum added by the asteroids therefore has to equal the amount $\delta \boldsymbol{L}_p$ the planet has to lose to maintain the balance; formerly, this means simply that

$$\delta \boldsymbol{L}_p = \Sigma \boldsymbol{L}_i$$

$$\delta ( m_p \boldsymbol{h}_p ) = \Sigma m_i \boldsymbol{h}_i . \qquad (16)$$

Using a simple scale analysis it can readily be shown that the magnitude of $\boldsymbol{h}_p$ is nearly constant to within four orders of magnitude and therefore the change in the planet's relative angular momentum required to maintain the needed balancing occurs almost entirely through its orientation. Accordingly, the magnitude can be moved outside the differential. Recalling that the $\boldsymbol{u}$ vectors are the normal unit vectors of the orbits under consideration, we then have that:

$$h_p \delta ( m_p \boldsymbol{u}_p ) = \Sigma m_i h_i \boldsymbol{u}_i . \qquad (17)$$

This expression states that the angular momenta of the asteroids is inducing a change in the planet's angular momentum which is caused primarily by the asteroids' mass and its spatial orientation relative to the plane of Planet X. The differential change in the total mass distribution affecting its angular momentum, therefore, has to equal the vectorial mass distribution of the asteroids as described by their mass and spatial orientation. Accordingly, we have that

$$h_p \Sigma m_i \boldsymbol{u}_i = \Sigma m_i h_i \boldsymbol{u}_i . \qquad (18)$$

By taking the dot product of both sides with the normal unit vector $\boldsymbol{u}_p$ and recalling that $\theta_i$ is the angle between the orbital planes of asteroid i and Planet X, we obtain

$$h_p \Sigma m_i \cos \theta_i = \Sigma m_i h_i \cos \theta_i .$$

And from this expression our estimate for $h_p$ can therefore be written as

$$h_p = ( \Sigma m_i h_i \cos \theta_i ) / ( \Sigma m_i \cos \theta_i ) ,$$

$$= \Sigma c_i h_i \cos \theta_i , \qquad (19)$$

where $c_i$ is the aggregate weighting factor for asteroid i, excluding the cosine factor in the numerator. The SRAM of Planet X is therefore equal to the weighted sum of the angular momenta of the asteroids. The corresponding vector form of (19) is then simply

$$\boldsymbol{h}_p = \Sigma c_i \boldsymbol{h}_i .$$

The sum of the angular momenta of the asteroids in (15) is a very small portion of those of the actual system as a whole as indicated above, but its role in equation (18) has an interesting implication nevertheless. When we take the partial derivatives of the sum on both sides of (18) with respect to the mass $m_i$ of a particular asteroid, the result implies that to a first order the specific relative angular momentum of the asteroids and Planet X are all equal ! As curious as this fact is, it was actually implicit in the equations of motion (6) above, restated below. As the torque of Planet X works its azimuth alignment effects on the asteroids, it also drives their relative angular momenta toward agreeing with its own in both orientation and magnitude as seen in the following equations of motion:

$$d\boldsymbol{h}_i / dt = \boldsymbol{\xi}_i \times ( \boldsymbol{h}_i - \boldsymbol{h}_p ), \qquad (20)$$

where $\boldsymbol{\xi}_i$ is the angular rate of change of the asteroid's relative angular momentum vector $\boldsymbol{h}_i$ about the nearly stationary $z$ axis of $\boldsymbol{L}_p$. The non zero difference in the relative angular momentum of any asteroid from that of Planet X, which is primarily due to the different orientation of their orbital planes, is gradually driving all such differences into the asteroid's specific relative angular momentum.

Although the caveats discussed in Section 2.0 obviously still apply, we find strong evidence of this effect having occurred in the asteroid data nevertheless. The average deviation of the relative angular momentum of the 12 asteroids from their mean value is only 11.7% of that mean. If we include the relative angular momentum of Planet X in the averages, the relative deviation goes up to only 12.6%. Clearly Planet X is driving the asteroids into an unusual form of resonance with its own relative angular momentum, as equation (20) suggests.

## 2.3 Eccentricity

The approach for estimating eccentricity leverages the angular locations of the two attractors discussed briefly in Section 2.1 above, which are arranging the asteroids' orientations along two angular directions of influence in azimuth. As argued above, the attractors are associated with the approach and retreat paths of Planet X, which produces a gravitational torque that gradually rotates each reference asteroid's major axis towards aligning with whichever attractor is closest to it in azimuth. There is an interesting equation from astrophysics that provides insight into this two-sided attraction. If we compute the average of the range measured from the orbital focus to a location on a Kepler orbit by averaging over all possible values of range, that average range value is well known to be the length of the semi major axis. Now, instead of averaging the orbital range by length, we instead calculate the average range ( there are of course two such ranges in the orbit ) over time, this is the center of the two angular directions where the body spends its time on average. Said another way, given a body's variable speed in orbit, the time-averaged range values are the two values of the body's range where the planet is located in its orbit on average in time. The following expression provides this time-averaged range as described in the literature (Plummer 1918):

$$\hat{r}_t = a ( 1 + \tfrac{1}{2} e^2 ) \qquad (21)$$

The two points on the planet's orbit having this value of range correspond to the two orbital locations on its inbound and outbound paths where the planet spends its time on average. These locations are the ultimate source of the double attractor in longitude of perihelion experienced by the asteroids on the opposite side of the Sun as the planet's torque drives their major axes towards aligning with one or the other associated average directions. The time-averaged range occurs when the planet is between aphelion and the termination points of the minor axis on the orbit, where the angular rate of the range is changing at a relatively low rate as we would expect. When we know the angular orientation from the Sun of either of these orbital locations, the associated range values can be derived from the well-known Kepler orbital equation:

$$\hat{r}_t = a ( 1 - e_p^2 ) / ( 1 + e_p \cos \alpha_p ), \qquad (22)$$

where $\alpha_p$ is the angle between PX's perihelion orientation and either of the two time-averaged range orientations, the cosine of either angle being the same. The angle is approximated from the estimated angular spread between the longitude of perihelion of PX and that of either of the two attractors, the lower one being used here corresponding to the planet's presumed approach path. Accordingly we have the following estimate for $\alpha_p$ as measured counter-clockwise from PX's perihelion:

$$\alpha_p = \Sigma_{\text{low}} ( 1/n ) ( \omega_i + \Omega_i ) - ( \omega_p + \Omega_p ) . \qquad (23)$$

By equating (21) and (22) we can solve for the required value of $\alpha_p$ associated with a particular eccentricity as

$$\cos \alpha_p = -3 e_p / ( 2 + e_p^2 ) . \qquad (24)$$

and using this expression to solve for $e_p$ and the statistical value of $\alpha_p$ from (23) above, the planet's eccentricity is approximated as

$$e_p \;=\; (\;-1.5 \;+\; (\,1.5^{\,2} \;-\; 2\cos^{2}\alpha_p\,)^{\,0.5}\;)\;\sec\alpha_p\,. \qquad (25)$$

From this equation it was found that using the 12 reference asteroids, the planet's eccentricity is about 0.65. Additional numerical results derived are provided and discussed in section 3.

## 2.4 Semi Major Axis, Period and Perihelion

The other orbital parameters are calculated directly from Kepler's laws using the eccentricity and semi major axis of PX estimated above. Given Newton's gravitational constant G and the mass of the Sun M , and the value of $h_p$ from equation (19), then the formal value of the semi major axis *a*, period *T* and perihelion *p*, respectively, are calculated as follows:

$$a \;=\; h_p^{\,2} \,/\, (\,G\,M\,(\,1 \;-\; e_p^{\,2}\,)\,)\,, \qquad (26)$$

$$T^{\,2} \;=\; \pi\;4^2\,a^3 /\; G\,M\,, \qquad (27)$$

$$p \;=\; a\,(\,1 -\; e\,)\,. \qquad (28)$$

The values obtained for *a*, *p* and *T* for the case of 12 reference asteroids were found to be 233 AU, 82 AU and 3556 years, respectively. Further results and discussion are provided in Section 3.

## 2.5 Mass of Planet X

When a closed system of bodies is moving solely under its own mutual gravitational attraction, the system's angular momentum is obviously conserved; less well known, however, is that the system's overall mass distribution takes on a certain specific order as well. Whenever we multiply the vector distance from the system's center of mass to each body of the system by the body's mass and add up all the resulting mass-weighted range vectors, the sum is always zero. This corollary to conservation of angular momentum is the key to estimating Planet X's mass. Because the Sun has over 99.9% of the mass of the inner solar system and because the approximation of Planet X's orbit provided above implies the planet's trajectory extends far outside the inner planets at its closest approach, our model presumes that we can take the inner solar system as a point mass centered at the Sun that includes all the orbital mass within Neptune's orbit, and that Planet X and the reference asteroids are the only remaining role players the model has to deal with. So if Planet X's mass is substantial, then its orbit around the Sun creates a reciprocal rotation of the Sun whereby the two bodies orbit about their common center of mass. The reference asteroids are presumed to have a collective mass several orders of magnitude smaller than that of Planet X, a fact we show below, and so their contribution to the system's mass distribution is negligible. Vectors in the following analysis have the Sun as the common point of origin.

Given these ground rules the system's mass-weighted range vectors relative to the system's center of mass can now be described. Let $\boldsymbol{R_p}$ be the vector distance between the Sun and Planet X, $\boldsymbol{r_{si}}$ the corresponding vector distance between asteroid *i* and the Sun, and *M*, $m_p$ and $m_i$ the mass of the Sun, Planet X and asteroid *i*, respectively. If we let $-\,\delta\,\boldsymbol{R_p}$ be the distance from the Sun to the system's center of mass and $\boldsymbol{R_p} - \delta\,\boldsymbol{R_p}$ the corresponding distance from the center of mass to Planet X, then we can formally write the mass balance equation for the system:

$$m_p\,(\,\boldsymbol{R_p} - \delta\,\boldsymbol{R_p}\,) \;-\; M\;\;\delta\,\boldsymbol{R_p} \;\;+\;\; \Sigma\, m_i\;\;(\,\boldsymbol{r_{si}} \;\;-\;\; \delta\,\boldsymbol{R_p}\,) \;=\; \boldsymbol{0}\,, \qquad (29)$$

where the three terms represent the mass /range distribution vectors of Planet X, the Sun and the reference asteroids, respectively. This expression can be somewhat simplified by first noting that

$$\delta\,\boldsymbol{R}_p \;\;=\; -\;\boldsymbol{R}_p\;\;m_p \,/\, (\,M \;+\; m_p\,) \;\;\cong\; -\boldsymbol{R}_p\;\;m_p\;\;/\,M\,, \qquad (30)$$

which is true because the mass of the Sun is at least four orders of magnitude bigger than that of Planet X. Secondly we can omit as negligible the second summation in the third term in (29), which is almost three orders of magnitude smaller than every other term. Using these approximations and some rearranging of terms, equation (29) becomes simply

$$m_p^2 \; \boldsymbol{R}_p \;=\; -\, M \, \Sigma \, m_i \; \boldsymbol{r}_{si} \,, \qquad (31)$$

where the minus sign indicates the range vectors are in opposite directions relative to the Sun. The two range vectors in (31) are obviously variables, so to complete the estimate we need to examine their average behavior over time. As explained in Section 2.2 changes in angular momentum of the system occur almost entirely within the collection of affected asteroids with their major axes symmetrically distributed over time about the planet's major axis, and leaving only minute changes to the angular momentum of the considerably more massive Planet X. As a result the mass-weighted range vector of Planet X behaves approximately as if there were no asteroids affecting its motion, and so the sum of the mass-weighted range vectors of the asteroids collective is nearly constant. That said we can make the problem more tractable if we project the asteroids' range vectors onto the planet's major axis by taking the dot product of the right hand side of (31) with $\boldsymbol{w}_p$, the unit vector pointing along that axis. Knowing that the average magnitude of the range vector in a Keplerian orbit is the length of the semi major axis, we can replace each resulting range magnitude in the sum with the value of its semi major axis, which results in a respectable approximation of the average value of the collective over time. For the left hand side of (31), taking the magnitude of $\boldsymbol{R}_p$ to be the length of its semi major axis, we also know that this specific range value occurs twice in an orbit, namely at the two points where the planet's minor axis terminates on its orbital path. The net average for the total is then computed by projecting the asteroid's average sum onto this axis using the angle between the two axes, the cosine of which is easily shown to be simply the eccentricity of PX. Collecting all these ideas together we have the final estimate as

$$m_p^2 \;=\; M \, \Sigma \, m_i \,( a_i / a_p )\, e_p \cos \varphi_i \,, \qquad (32)$$

where $\varphi_i$ as indicated above is the angle between the major axes of PX and asteroid i, $a_i$ and $a_p$ are lengths of the semi major axes of a particular asteroid and PX, respectively, and $e_p$ is the planet's eccentricity. The associated diameter of PX was computed assuming the planet has the same density as that of the Earth. Notice that with the mass of Planet X now determined, equation (31) provides a direct way of approximating the current location of the planet.

The estimate of the mass of Planet X increases as the square root of the collective mass of all bodies under its clustering influence, presumed here to be the 12 reference asteroids used in the study. Inasmuch as the mass of the reference asteroids in general is not well known nor is the total number of asteroids similarly affected, the mass estimate is more uncertain than that of the other parameters due to the limitations of the available data. In fact the particular asteroids being used are only known to astronomy because their long period orbits have brought them within observational range only over the last few decades. Using equation (32) and all 12 asteroids for the calculation, the mass of PX was found to equal approximately 8.4 Earth masses with a diameter of about 25900 km.

## 2.6 Proof-of-Concept Orbital Simulation

A limited-scope, proof-of-concept simulation was developed to verify that the general behavior is at least partially as predicted qualitatively for the case considered. The orbits of selected bodies were integrated out to 40 million years with the Lie-integrator (Hanslmeier & Dvorak 1984; Eggl & Dvorak 2010; Galiazzo, Bazso & Dvorak 2014) at precision -13 using the limited solar system described below. The simulation examined the collective behavior of four bodies ( the Sun, Planet X, and the two dwarf planets 90377 Sedna and 2012 VP113, with the orbital elements of the last two taken at the same epoch, 12/25/2016 00:00 UT) over a 40 million year time span. The orbital configurations and mass of the bodies and other related related parameters were those presented in Table 1, with those of Planet X taken from the estimated values using all 12 asteroids. The two dwarf planets were selected for the run primarily because the orbit of one, Sedna, is inclined below the orbit of Planet X and that of 2019 VP113 is inclined well above it, which allowed exercising the theory for the two primary cases of interest discussed in Section 2.0.

The graph in Figure 2 below depicts the behavior of five key parameters of the run: the inclination of the three bodies of interest over the time span and the angular difference $\theta_i$ between the orbital plane of Planet X and that of either dwarf planet, the calculation of that angle being defined in equation (11). As predicted the inclinations of both bodies steadily moved towards that of Planet X, although their difference angles did not converge in the same time frame. For asteroids with longitudes of the ascending node close to that of Planet X, the difference angle is almost totally comprised of inclination and the convergence of the two planes is relatively uncomplicated and in close synchrony with the convergence of inclination. This is not at all the case with the two dwarf planets, however, since the planet's longitude of the ascending node deviates from that of Sedna's by over 35° and that of 2012 VP113 by over 17° in the opposite direction. In both cases, but in particular with Sedna, the large longitudinal deviation was seen to drive the difference angle to a higher value as we observed the beginnings of an extended period of oscillation to reach full convergence occurs. Notice too that "convergence" does not imply that an asteroid will have the same inclination and longitude of the ascending node as that of Planet X, only that the planar difference angle is small. This fact is very clear from the orbit of Sedna itself whose difference angle is only about 9° and yet as noted its longitude of the ascending node deviates from that of Planet X by over 35°. This simulated behavior provides qualitative support although by no means definitive confirmation for the complete dynamic behavior predicted.

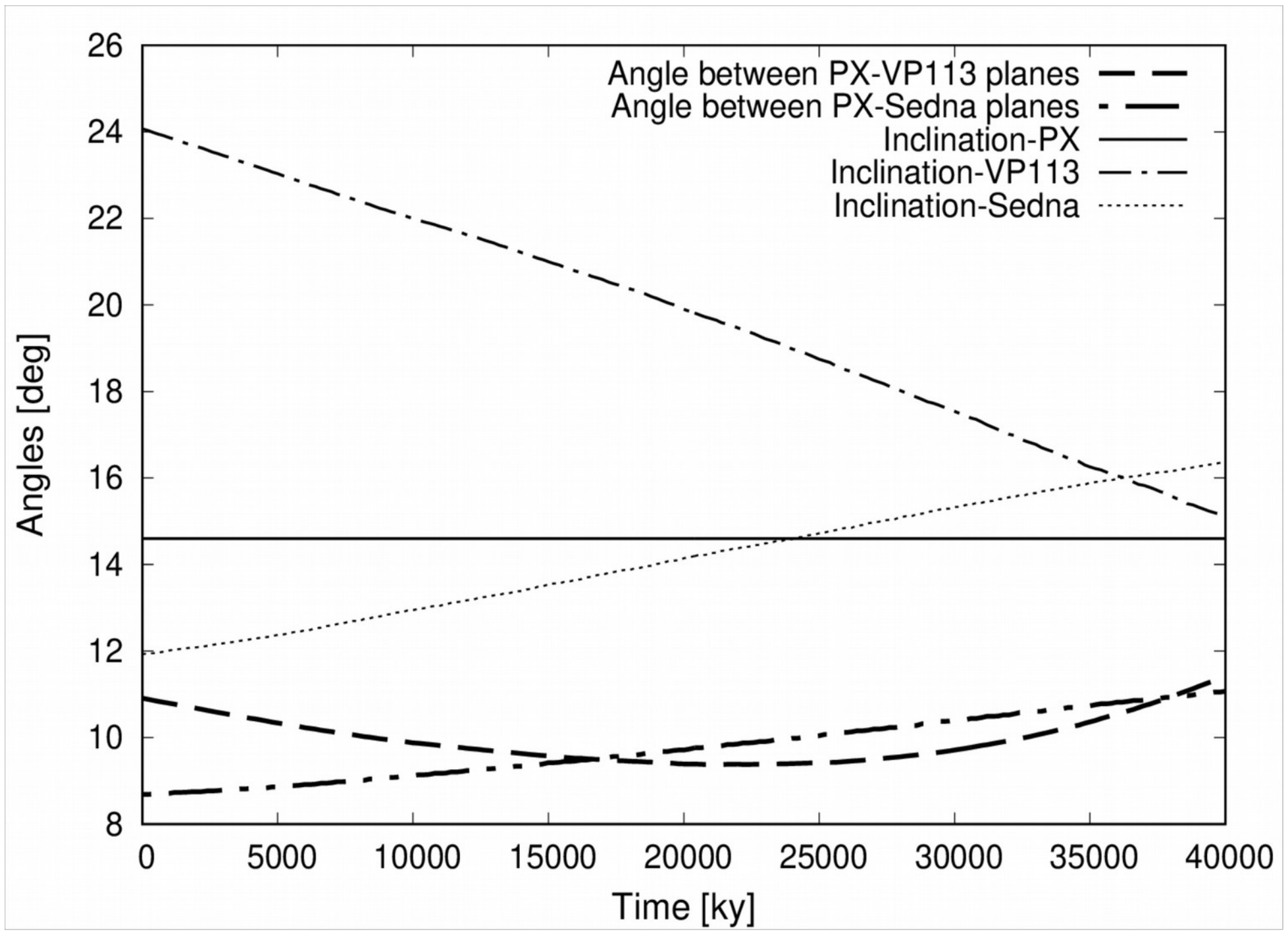

Fig 2 Inclinations of Planet X, 90377 Sedna and 2012 VP113 and angular deviations between Planet X's orbital plane and those of the other bodies

## 3. Results and Discussion

Numerical results of the analysis are summarized in Table 1 below with estimates of Planet X's orbital parameters given in the last rows and angular comparisons of orbital characteristics of Planet X with the reference asteroids in the last two columns. All parameters of PX were estimated using the mathematical models presented herein for both the 6 asteroids from Brown's original paper and 12 asteroids – the original 6 being the first 6 objects listed in the table, plus 6 additional asteroids selected by the authors. The density values used have been taken from the those estimated for trans-Neptunian objects and centaurs from Johnston's Archive (Johnston 2019), with values ranging from 0.46 to 2.64 g/cm$^3$, which provided the data source for the densities used to estimate the asteroids' mass values and relative angular momentum. Accordingly, for objects with diameters greater than 900 km, the mean densities were presumed to be 1.96 g/cm$^3$. For objects with diameters less than 900 km but greater than 400 km, the mean densities were presumed to be 1.06 g/cm$^3$, and for objects with diameters less than 400 km the presumption was 0.94 g/cm$^3$.

The working hypothesis of this analysis is that Planet X is inducing gravitational torque on the asteroids, whose primary effect is to drive their orbital planes toward aligning with its own. If this is true then the asteroid's orbital planes themselves provide the key to determining Planet X's orbital plane. Using this premise, geometric considerations and statistical estimation were used to generate an estimate of the planet's orbital plane that best fits the corresponding orbital planes of the asteroids. From the analytical description of this best fit plane, the planet's inclination and longitude of the ascending node were computed directly and uniquely. Because the description of this estimated plane is unique given the analytical correlations involved, the two associated coordinates are also unique and fixed for all further estimates involving the planet's orbit. The resulting estimate was found to agree very well with the reference data; the average angular difference in magnitude between the estimated orientation of this plane and those of all 12 asteroids is only -2.4°, and the sample standard deviation is 14.7°. Applying the standard confidence interval criteria to the estimate, we have with 95% confidence the planet's true angular orientation falls within about ± 8° of that estimated value. The data therefore clearly implies that this plane is highly likely to be the orbital plane of the large body perturbing the asteroids' orbits. Column 13 of the table provides the angular deviations between this plane and those of the 12 reference asteroids.

**Table 1** Source reference asteroid orbital data and associated estimates of Planet X's orbital parameters and mass (last two rows)

| Object | Period ( yrs ) | *a* (AU) | *e* | *p* ( AU ) | *ι* ( Deg ) | *Ω* ( Deg ) | *ω* ( Deg ) | Diameter ( km ) | Mass ( kg ) | *h* ( m*2/sec ) | *ϖ* ( Deg ) | *θ* (Deg) 12 Obj | Δ *ϖ* (Deg) 12 Obj |
|---|---|---|---|---|---|---|---|---|---|---|---|---|---|
| 2012 VP113 | 4136 | 257 | 0.69 | 80 | 24.0 | 90.8 | 293.8 | 702 | 1.920 e20 | 5.179e10 | 384.6 | +10.9 | +5.6 |
| 2013 RF98 | 5600 | 349 | 0.88 | 36 | 29.6 | 67.6 | 316.5 | 55 | 8.210e16 | 3.681 e10 | 384.1 | +20.5 | +5.1 |
| 2004 VN112 | 5850 | 315 | 0.85 | 47 | 25.6 | 66.0 | 327.0 | 152 | 1.728 e18 | 4.155 e10 | 393.0 | +17.4 | +14.0 |
| 2007 TG422 | 12789 | 472 | 0.93 | 36 | 18.6 | 113.0 | 285.8 | 174 | 5.384 e18 | 3.708 e10 | 398.8 | +4.3 | +19.8 |
| 90377 Sedna | 11400 | 506 | 0.85 | 76 | 11.9 | 144.5 | 311.5 | 995 | 1.026 e21 | 5.283 e10 | 456.0 | -8.7 | +5.0 |
| 2010 GB174 | 6945 | 364 | 0.87 | 48 | 21.5 | 130.7 | 347.7 | 250 | 7.688 e18 | 4.219 e10 | 478.4 | +9.8 | +27.4 |
| 2000 CR105 | 3442 | 223 | 0.80 | 44 | 22.7 | 128.3 | 317.0 | 285 | 1.127 e19 | 3.968 e10 | 445.3 | +10.3 | -5.7 |
| 2014 SR349 | 4960 | 291 | 0.84 | 48 | 18.0 | 34.8 | 341.5 | 200 | 3.936 e18 | 4.182 e10 | 376.3 | -19.4 | -2.7 |
| 2010 VZ98 | 1846 | 150 | 0.77 | 34 | 4.5 | 117.4 | 313.9 | 401 | 3.55 e19 | 3.460 e10 | 431.3 | -10.2 | -19.7 |
| 2003 SS422 | 2744 | 196 | 0.80 | 39 | 16.8 | 151.1 | 210.8 | 168 | 2.333 e18 | 3.734 e10 | 361.9 | -11.7 | -17.1 |
| 2000 OO67 | 11500 | 510 | 0.96 | 21 | 20.1 | 142.3 | 212.2 | 44 | 4.191 e16 | 2.858 e10 | 354.5 | -11.4 | -24.5 |
| 2013 UT15 | 2740 | 196 | 0.78 | 44 | 10.7 | 191.9 | 252.3 | 274 | 1.012 e19 | 3.938 e10 | 444.2 | +17.1 | -6.8 |
| **Planet X (6)** | 3375 | 225 | 0.62 | 86 | 19.5 | 95.5 | 153.5 | 26400 | 5.388 e25 | 5.239 e10 | 249.0 | N/A | N/A |
| **Planet X (12)** | 3556 | 233 | 0.65 | 82 | 14.6 | 107.7 | 127.3 | 25900 | 5.020 e25 | 5.168 e10 | 235.0 | N/A | N/A |

The secondary effect on the asteroids' orbits involves a much smaller component of the induced torque, which drives the major axis of each reference asteroid in azimuth, i.e. rotates the orbit about an axis normal to the planet's orbit located at its solar focus. This effect tries to align the asteroid's major axis with either the approach or departure path of Planet X, the one "selected" being the one closest in azimuth to the asteroid's major axis. The angular difference for each asteroid between its major axis and its particular path of "attraction" (i.e. approach or retreat path) are given in the last column of the table. ( Note that Planet X's azimuth given in column 12 uses the estimated value of the longitude of perihelion plus 180° so that comparison with those of the asteroids is more apparent.) The double clustering is clearly evident in the data, in that the azimuth orientation, i.e. the longitude of perihelion, of each of the 12 reference asteroids clearly fall into two distinct clusters: a lower-valued cluster in the interval of about 355° to 400° and a second of about 430° to 480°. The two clusters are separate distributions whose mean values have a spread of about 72° in the 12 asteroid case, implying a midpoint of about 415° axis of symmetry between the two distributions, formally 235° taking this direction to be the Planet X' longitude of perihelion, where the 180° reduction is due to the planet's perihelion being in the opposite direction as that of the reference asteroids. And knowing this value we calculated the final coordinate of PX, the argument of perihelion, and having all three coordinates the planet's perihelion was located at right ascension 236.32° and declension –7.98° placing it in the constellation Libra. By adopting this orientation as the major axis of Planet X's orbit and estimating the azimuth of its approach and departure paths as described in Section 2.1, PX's azimuth orientation was compared with those of the reference asteroids. This was done for all 12 asteroids by comparing the longitudes of perihelion of the asteroids in the larger and smaller azimuth grouping with the argument of perihelion of PX's approach path and departure direction, respectively. Again we have good agreement between the two with a sample standard deviation of 15.9° for the case of the 12 asteroids. By combining the two identical distributions, we computed a 95% confidence interval of ± 8.9° for the attractors' longitude of perihelion.

An estimate for PX's specific relative angular momentum (SRAM) was developed using perturbation analysis involving the angular momentum of the entire system, from which it was shown that the planet's SRAM relates directly to the angular momenta of all the reference asteroids and inversely to their collective mass. The higher an asteroid"s angular momentum, the greater its contribution to the estimation and vice versa, with very small objects playing little role in the estimation. As seen in the table the estimate produced is very close to those of the heaviest reference asteroid, Sedna, whose mass strongly dominates the collective. A particularly striking aspect of the results is that the estimated SRAM for Planet X and that of the known values for the reference asteroids are all very close to the same value. In fact the average deviation of the relative angular momentum of the 12 asteroids from their mean value was found to be only 11.7% of that mean value. If we include the relative angular momentum of Planet X in the averages, the relative deviation goes up to only 12.6%. This highly counter intuitive result was shown to follow directly from the equations of motion derived; gravitational torque is driving the asteroids' relative angular momenta to align with that of PX as it shifts the orbits' orientations spatially.

The eccentricity was estimated by analyzing PX's approach and departure paths in relation to the angular orientations of the reference asteroids' major axes. Because the orientations of the these axes cluster around two symmetric angular deviations from PX's major axis, it was concluded that the two average separations of these directions should correspond to the two points where PX spends its average time in orbit. This claim is strongly supported by the data indicated in the above discussion. This preferred angle allowed calculating the eccentricity of PX directly, which turned out to be about 0.65. The calculation is dependent only on the estimate of PX's longitude of perihelion, and therefore the eccentricity's confidence interval is derivative of that angle. Accordingly, using this fact solving gives an approximate 95% confidence interval for the eccentricity of $\pm 0.11$ : specifically $.54 \le e_p \le .76$.

Having the eccentricity then allowed using the Kepler's orbital equations to calculate other elements of PX's orbit, which for the 12 asteroid case were found to be 233 AU for the semi major axis, 3556 years for the period and 82 AU for perihelion. For the 6 asteroid case, the reader is referred to the table below for the corresponding estimates. The interesting thing about these numbers is that they differ substantially from results proposed by the original Caltech team and virtually all subsequent publications.

Estimation of PX's mass was shown to follow directly from arguments involving conservation of angular momentum of the entire system and standard probability theory. Using these ideas the mass of PX was found to be proportional to the square root of the asteroids' collective mass, adjusted for geometric considerations. This relationship is fortuitous in that errors in estimates of the asteroids' collective mass are modulated significantly. If the actual collective mass is 100% in error, or twice the actual value, then this error is reflected in the planet's mass estimate as only a 40% error. The final resulting estimate of PX was found to be 8.4 Earth masses, similar to estimates found in recent literature.

The estimated orbit of Planet X has revealed several unexpected surprises on final analysis. The first of these, discussed above, is that the planet is apparently driving the asteroids' specific relative angular momentum vectors towards its own in an interesting form of geometric resonance. More surprising yet is the number of bodies in or near mean motion resonance with the planet, four of which are among the 12 reference asteroids themselves: Sedna with a period of 11400 years ( 1:3 ), 2013 RF98 with a period of 5600 years ( 2:3 ), 2000 OO67 with a period of 11500 years (1:3), and 2010 VZ98 with a period of 1846 years ( 2:1 ). Other likely resonances include three long period Trans Neptunian objects in elliptical orbits: 1999 DP8 having a moderately inclined elliptical orbit with a period of 1250 years ( 3:1 ), 2010 QQ7 with a smaller inclination and a period of about 1233 years ( 3:1 ), and 2015 VD168 with a small inclination and a period again of about 1233 years (3:1). Interestingly, the current positions of these seven bodies along with the planet's estimated orbit provide sufficient data to enable another method of estimating the current position (right ascension and declination) of PX, the first method being noted at the end of Section 2.5.

Finally, from its orbit we can compute approximatively its apparent magnitude in visual band at certain positions from this equation which is described in the IAU H-G two-parameter model described in "Asteroids II" (Univ of Arizona Press, 1989), p 549-554: $h_v = H + 5 \log10 \delta + 5 \log10 r - 2.5 \log10 ((1-G)\Phi_1 + G\Phi_2)$ δ is distance from Earth, 'r' distance from Sun, 'H' is measured absolute magnitude, 'G' is slope parameter assumed 0.15 if not measured, $\Phi_1$ and $\Phi_2$ are phase coefficients. B= tan( 0.5 * phase_angle ) $\Phi_1$ = e-3.33 B^0.63 , $\Phi_2$ = e -1.87 B^1.22 . Now supposing the planet is at its perihelion or nearly, r=84.4 au(from our computation q=84.35 au), with a favorable phase angle of 10°, we have an heliocentric distance of 83.4 au (Theorem of Carnot), thus the apparent magnitude of PX would be 18.93 mag, assuming that H=-1.1 like 136199 Eris, or if we assume Sedna's absolute magnitude, it would be 21.53 mag. If the object is far away between the perihelion and aphelion (397.65 au), so at about 241 au apparent magnitudes in visual with the previous properties range between 23.50 mag and 26.10 mag, and this result explains why it is so difficult to observe a planet at such distance, probably with a magnitude of ~24.8 at a favorable phase angle, which only very large telescope might observe it.

## 4. Conclusions

The theory that a large, remote planet orbiting the Sun is systematically arranging long period asteroids is strongly supported by the analysis and results described herein. The approach has celestial mechanic, applied engineering physics and standard statistics to analyze the planetary system's angular momentum and gravitational torque in the underlying dynamics that enabled the orbit and mass of the unseen planet to be estimated and the orderly arrangement of the 12 reference asteroids to be understood mathematically. All results presented were derived from these math models. Although the analysis began with the original set of 6 objects proposed by Caltech astronomers (Batygin & Brown 2016), 6 additional objects selected by the authors were also analyzed together with the original 6 to reduce the statistical uncertainties associated with the limited size of the original data set and to improve the overall fidelity of the estimates.

The foundational concept of the approach has been that the pattern and geometry of the asteroids' orbits can be used to infer the orbit of Planet X statistically. Primarily three properties of the asteroids were used to develop the solution: (1) the spatial orientation of the asteroids' orbital planes and major axes; (2) the specific relative angular momentum of the asteroids; and (3) the size and spatial distribution of the asteroids' mass. Using the physics of orbital mechanics and standard statistics applied to the orbits and mass of the reference asteroids, the two data sets were used to develop estimates of the orbital characteristics of Planet X from first principles that included the planet's inclination, longitude of the ascending node, argument of perihelion, specific relative angular momentum, eccentricity, semi major axis length, orbital period, perihelion and mass.

The most dramatic result discovered was that the orbital planes of all 12 reference asteroids fall quite close to a common plane, predicted by the theory to be the orbital plane of Planet X. The angular separations between the asteroids' orbital planes and that of the central plane, with an estimated inclination of 14.6°, and longitude of the ascending node of a little under 108°, was found to be only 12.6°, on average, 14.1°, in standard deviation and a 95% confidence interval of ± 8.0°. Similarly, the azimuth orientations of the asteroids' major axes were found to align as predicted in two distinct directions separated by about 72°, and consistent with the proposed gravitational torque predicted to produce two regions of attraction in azimuth in the directions of the approach and departure paths of a massive body in a wide orbit with a long period. From this result the argument of perihelion was found to be a little over 127°, which places the planet's perihelion in Libra. The magnitude of the angular deviations of these axes from the average azimuth orientations of the two near-solar orbital paths calculated from the planet's estimated orbit developed herein proved to be only 13.2°, with a corresponding sample standard deviation of 15.9°, and 95% confidence interval of ± 9.2°. again good confirmation of the effect of gravitational torque proposed. The reference coordinates calculated from these orientations using only the original 6 objects agree closely with those proposed at Caltech (Batygin & Brown 2016), providing additional confirmation of our approach. Using the larger sample of 12 clearly showed that the argument of perihelion is actually substantially smaller when a more statistically representative sample is used in the estimation.

Using perturbation analysis applied to the system's total angular momentum, the specific relative angular momentum of PX was shown to equate to a weighted sum of the specific relative angular momenta of the reference asteroids. From this relationship and from the orbital motion of the reference asteroids it was shown that the specific relative angular momenta of all the reference asteroids are being driven to coincide with that of Planet X. This unusual prediction turns out to be confirmed by the data, which provides support that the model is reflecting what is actually happening regarding angular momentum. The average deviation of the relative angular momentum of the 12 asteroids from their mean value was found to be only 11.7%. If we include the relative angular momentum estimated for Planet X in the averages, the relative deviation goes up to only 12.6%.

Estimation of PX's mass was shown to follow directly from classical arguments involving conservation of angular momentum of the system and the average behavior of the clustered asteroids. The planet's mass was found to be proportional to the square root of the asteroids' collective mass, moderately adjusted for geometric considerations. The resulting estimate using all 12 asteroids of about 8.4 Earth masses is consistent with recent estimates found in the literature (Batygin & Brown 2019).

Eccentricity was derived using the planet's dominant angular directions of approach and retreat approximated from the asteroids' azimuth clustering used to estimate the argument of perihelion, an approach which is relatively insensitive to the reference asteroids used in the estimation sample. From this result the eccentricity was estimated to be about 0.62 for the case of 6 asteroids and 0.65 for the 12, close to the original estimate estimate of 0.60 (Batygin and Brown 2016). Using the value of 0.65, the SRAM value and orbital coordinates described above, the major axis length, orbital period, and perihelion were calculated directly from Kepler's equations. The values obtained were found to be smaller than current predictions at 233 AU, 3556 years, and 82 AU, respectively. The particular period and large mass estimated are apparently responsible for possibly seven mean motion resonances among the reference asteroids and other TNOs. Another form of resonance observed is that the specific relative angular momentum of all clustered bodies seem to be driven towards that of Planet X, a particularly interesting effect in that the full angular momenta of these bodies differ by several orders of magnitude.

A limited proof-of-concept simulation was generated to ensure the integrity of the overall model and orbit predicted and to verify qualitatively that the general behavior is as expected. The simulation integrated the orbits of Planet X, and the dwarf planets Sedna and 2012 VP113 out to 40 million years, from which the approximate convergence of the inclinations of the dwarf planets was observed, although full convergence of the orbital planes was still in the distant simulation's future.

Final analysis indicated that Planet X is the cause of possibly seven near mean motion resonances, four of which are in the set of 12 reference asteroids analyzed. It was noted that the large number and known current position of the bodies in resonance allow deriving an explicit estimate of the current position (right ascension and declination ) of Planet X.

The analysis and results described herein provide compelling evidence for the existence of a large planetary body in a long-period orbit with the characteristics predicted. Evidence from 12 reference asteroids show a predictable pattern of behavior and strong statistical correlation regarding their orbital planes, azimuth orientations, mass and specific relative angular momenta, consistent with the proposed orbital characteristics and mass of a large planet in a long elliptical orbit.

As for a possible observation, Planet X, would range between V=18.9 and 26.1, probably with a magnitude of ~24.8 at a favorable phase angle, which only very large telescope might observe it.

## References

1. Batygin, K., & Brown, M. 2016, AJ, 151, 22
2. Batygin, K., & Brown, M. et al 2019, ISO, NLM 805, 1
3. Brunini A. 1992, Celest. Mech. Dyn. Astron. 53, 129
4. Collander-Brown S., Maran M. & Williams I.P. 2000, 318, 101
5. de la Fuente Marcos, C. & de la Fuente Marcos, R. 2014, MNRAS 443, L59
6. Eggl, S. & Dvorak, R., in "Dynamics of Small Solar System Bodies and Exoplanets", eds Souchay J. & Dvorak, R., Springer: Berlin, Heidelberg p. 431
7. Galiazzo, M., Bazso, A. & Dvorak, R. 2014, Planet Space Sci., 84, 5
8. Gomes R., Matese J. & Lissauer J. 2006, Icarus 184, 589
9. Gomes, R., Deienno, R. & Morbidelli, A. 2016, AJ 153(1)
10. Gunn E.J. 1970, New Sci, 48, 345
11. Hanslmeier, A. & Dvorak, R., 1984, A&A 132, 203
12. Harrington, R. 1988, AJ, 96, 1476
13. Hildebrand, F. 1962, Advanced Calculus for Applications, Prentice Hall: Englewood Cliffs
14. Johnston, W. 2018, in: TNO/Centaur diameters, albedos, and densities. Johnston's Archive. Available via PDS small bodies node. http://www.johnstonsarchive.net/astro/tnodiam.html, title of subordinate document. Cited 3 July, 2019
15. Lykawka P.S., & Mukai T. 2008, AJ, 135, 1161
16. Malhotra, R., Volk, K. & Wang, X. 2016, ApJ, 824, L17
17. Matese J.J., Whitmire D.P. & Lissauer J.J. 2005, Earth Moon Planets 97, 459
18. Melita M.D. & Williams I.P. 2003, Earth Moon Planets 92, 447
19. Melita M.D., Williams I.P. et al 2004, Icarus 171, 516
20. Moulton, F.R. 1984, "An Introduction to Celestial Mechanics", Dover: New York
21. Plummer, H. C. 1918, "An Introductory Treatise of Dynamical Astronomy", Cambridge University Press: London, p. 38
22. Trujillo, C. & Sheppard, S. 2014, Nature, 507, 471